\documentclass[conference]{IEEEtran}
\ifCLASSINFOpdf
  \usepackage[pdftex]{graphicx}
\else
   \usepackage[dvips]{graphicx}
\fi
\usepackage[cmex10]{amsmath}
\usepackage{amssymb}
\usepackage{algorithm,algorithmic}
\usepackage[font=footnotesize]{subfig}
\usepackage[applemac]{inputenc}

\begin{document}
%
\title{Selecting source image sensor nodes based on 2-hop information to improve image transmissions to mobile robot sinks in search \& rescue operations}

\author{
\IEEEauthorblockN{Congduc Pham}
\IEEEauthorblockA{LIUPPA, University of Pau\\
Pau, France 64013\\
Email: congduc.pham@univ-pau.fr}
\and
\IEEEauthorblockN{Mamour Diop}
\IEEEauthorblockA{LIUPPA, University of Pau\\
LANI, Gaston Berger University\\
Email: mamour.diop@univ-pau.fr}
\and
\IEEEauthorblockN{Ousmane Thiaré}
\IEEEauthorblockA{LANI, Gaston Berger University\\
Saint-Louis, Senegal\\
Email: ousmane.thiare@edu.ugb.sn}}


%



\maketitle

\begin{abstract}
We consider Robot-assisted Search\&Rescue operations enhanced with some fixed image sensor nodes capable of capturing and sending visual information to a robot sink. In order to increase the performance of image transfer from image sensor nodes to the robot sinks we propose a 2-hop neighborhood information-based cover set selection to determine the most relevant image sensor nodes to activate. Then, in order to be consistent with our proposed approach, a multi-path extension of Greedy Perimeter Stateless Routing (called T-GPSR) wherein routing decisions are also based on 2-hop neighborhood information is proposed. Simulation results show that our proposal reduces packet losses, enabling fast packet delivery and higher visual quality of received images at the robot sink.
\end{abstract}


%
\IEEEpeerreviewmaketitle

\section{Introduction}
\label{sec:intro}
Robot-assisted Search\&Rescue is an important domain of research and many contributions on specific robot hardware and control software have been proposed by the scientific community. In some cases, robots are designed to have a certain degree of autonomy and therefore can move and explore by themselves and provide information to a control center which is usually under the supervision of a human operator. The advantage of robots is clearly in their ability to evolve in very dangerous areas where human rescue workers can hardly go. One main issue however is the limited operating time of battery-operated autonomous robots and it is essential to optimize the robot exploration process. In this paper, we consider Robot-assisted Search\&Rescue operations enhanced with some fixed image sensor nodes capable of capturing and sending visual information either on demand or on relevant event detection. These image sensors can be thrown in mass to provide visual information in various geographical parts of an area of interest.

\begin{figure}[H]
\centering
\includegraphics[width=\linewidth]{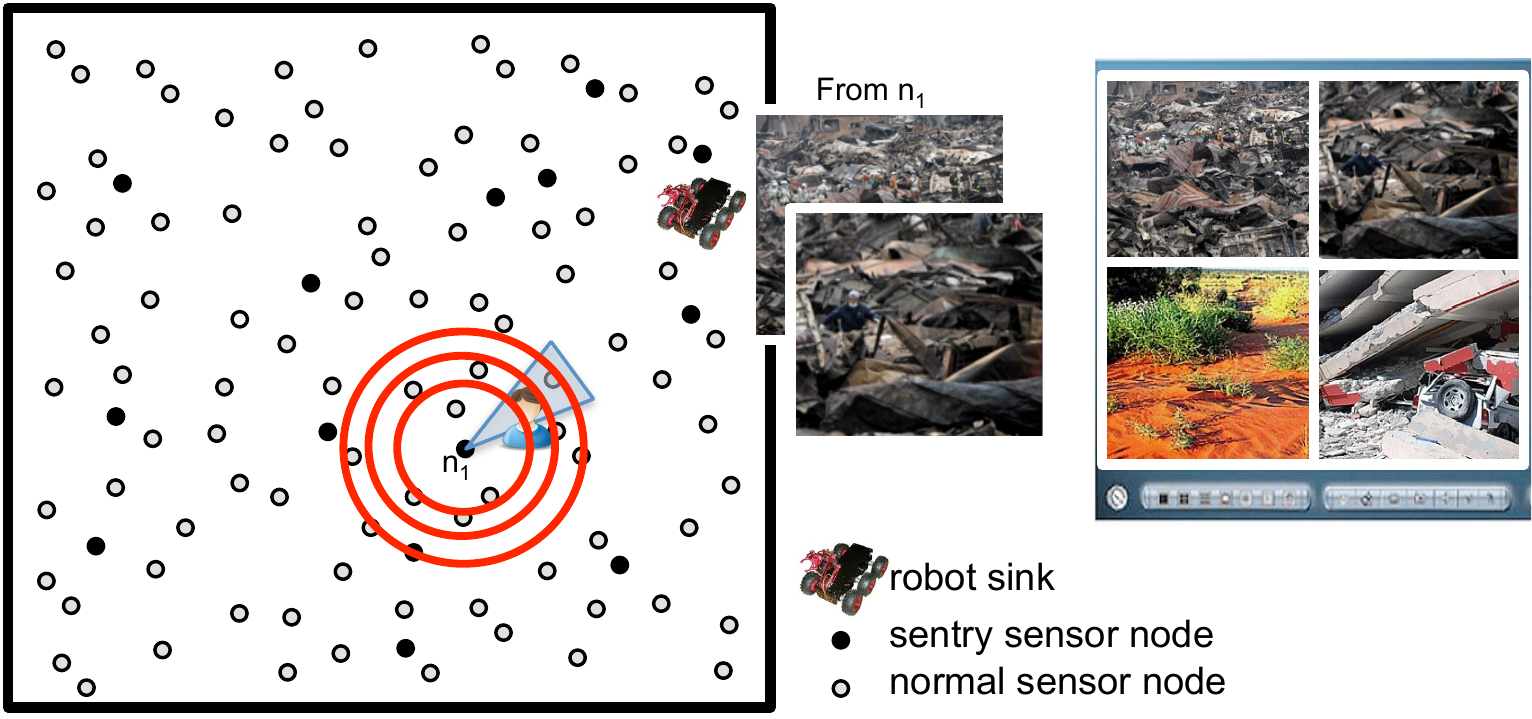}
\caption{Fixed image sensors and mobile robot sink}
\label{intrusion}
\end{figure}

Figure~\ref{intrusion} shows the scenario of a random deployment of image sensor nodes. The deployed image sensor network could be used for situation awareness in Search\&Rescue applications for instance where images from remote nodes are collected at a control center, displayed and possibly integrated into a GIS system. Such information can be used to optimize the deployment of robots, taking into account the terrain condition for selecting the most appropriate robot technology for instance. Once robots are deployed and are moving in the field, deployed image sensor nodes can be used to provide visual information of remote areas to the robots. Robots can then use these images to perform advanced image processing tasks either for detecting events of interest (such as a human victim presence) or for optimizing their exploration paths. In addition, when these autonomous robots need to send images back to the control center for the human operator the deployed sensor network can be used as a relay network if necessary.

\begin{figure}[htb]
\begin{center}
\includegraphics[width=0.8\linewidth]{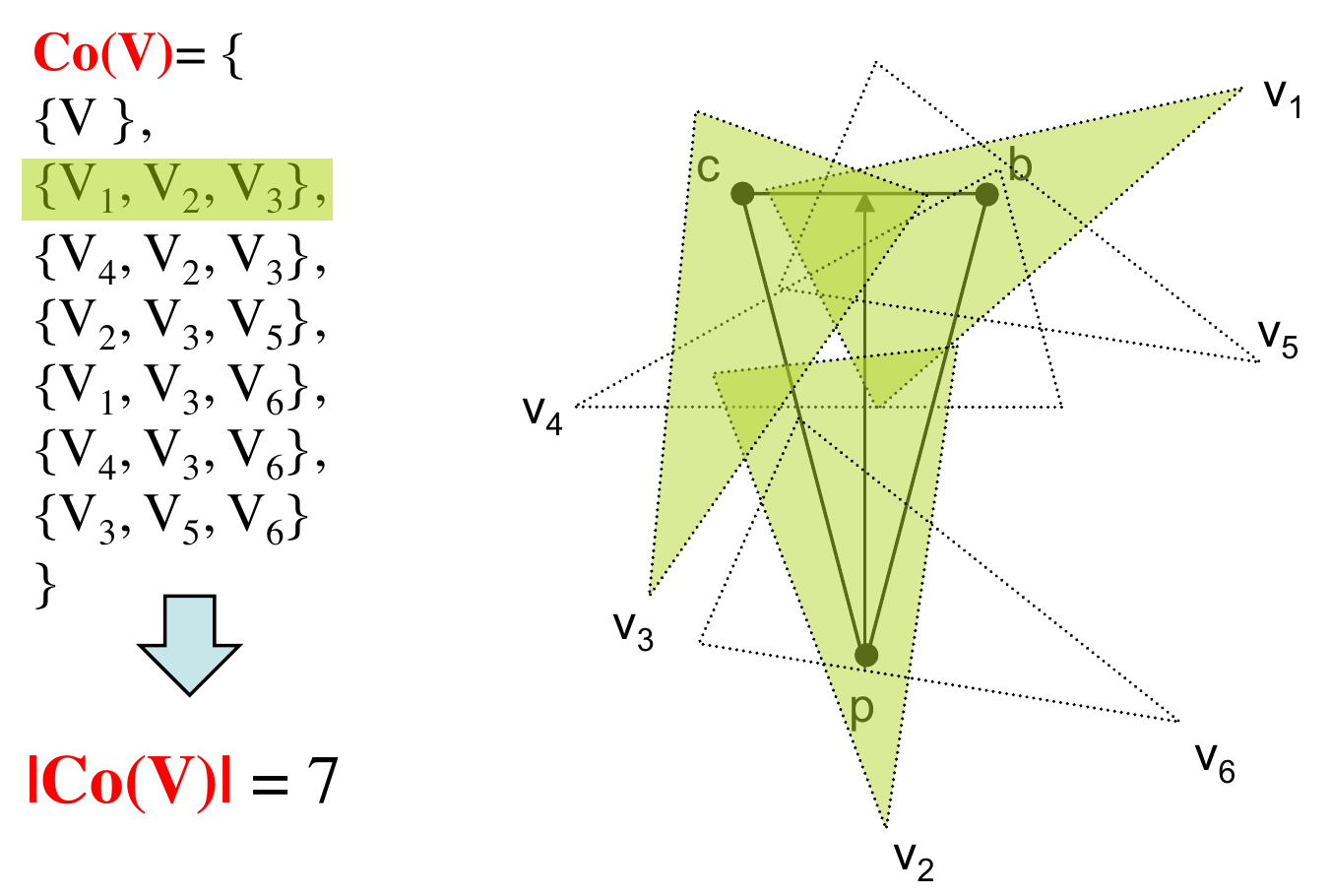}
\caption{Coverage model and cover set.}
\label{fig:coverset}
\end{center}
\end{figure}

In our previous work on image sensor networks, we considered that sensor nodes can be redundant (nodes that monitor the same region) leading to overlaps among the monitored areas \cite{Mak10}. In figure \ref{fig:coverset}, the Field of View (FoV) of a sensor $V$ is represented by the triangle $(pbc)$. A cover set for $V$ is defined as a subset of image nodes which covers its FoV area. If we consider nodes $V$, $V_1$, $V_2$ and $V_3$ the possible set of cover sets is $Co(V) = \Big\{ \{V\}, \{V_1, V_2, V_3\} \Big\}$.  If we add nodes $V_4$, $V_5$ and $V_6$, $Co(V)$ has more elements as depicted in figure \ref{fig:coverset}. One commonly found approach for sensor node activity scheduling consists in putting in sleep mode nodes whose sensing area are covered by others. However, in mission-critical applications where responsiveness must be increased, nodes that possess a high redundancy level (their sensing area are covered many times by other nodes so that they have many cover sets) could rather be more active than other nodes with less redundancy level. In \cite{Mak10,Pham11a} the idea we developed is that when a node has several covers, it can increase its frame capture rate to act as a sentry node because if it runs out of energy it can be replaced by one of its cover sets. In Figure~\ref{intrusion}, sensor nodes can organize themselves to designate a number of sentries (nodes in black) to better detect events, send images and to possibly trigger alerts.

In the scenario we consider here, deployed image sensor nodes can have very basic event detection capabilities and mobile robots have more powerful processing capabilities to act as a Sink for the fixed image sensor nodes. When an image node detects an event, it will \textit{(a)} send one or several images to the mobile robot sink and \textit{(b)} activate one of its cover sets. On activation, cover set members will also send one or several images to the robot sink to provide more information for disambiguation purposes. We can see that an event detection triggers the simultaneous transmission of a large volume of visual data from multiple sources to the robot sink: even with an optimized image encoding scheme, a 320x320 pixels image in 256 gray levels needs about 16KB of data. Given the limited performance level of small autonomous image sensor nodes, in terms of processing capability, radio technologies and transmission bandwidth \cite{Pha13-wimob}, with no control, the transmission of many images produces significant data losses due to network congestion, therefore degrading the visual information quality at the robot sink. Obviously, cover sets of a given node have different size, level of coverage/energy, and, most importantly, different performance levels for transferring large amount of data to the robot sink. In the context of mission-critical application, detecting events is important but receiving high quality images at the lowest latency is also very important. Our objective in this article is to significantly reduce congestion and increase image quality at the robot sink when simultaneous images are sent from image sensor nodes. We will present our approach based on 2-hop neighborhood knowledge to optimally select the most appropriate cover-set for image transmission. Then, in order to be consistent with our proposed approach, a multi-path extension of Greedy Perimeter Stateless Routing (called T-GPSR) wherein routing decisions are also based on 2-hop neighborhood information is also presented. 

The paper is structured as follows. Section \ref{sec:relWork} outlines related 2-hop information-based algorithms. Section \ref{sec:selection} describes the main guidelines of the proposed cover set selection approach for efficient image transmission. T-GPSR, our 2-hop information-based GPSR extension,  is then presented in Section \ref{sec:tgpsr}. Simulations and results are shown in Section  \ref{sec:results} and we conclude in Section \ref{sec:conclusion}.

\section{2-hop neighborhood information}
\label{sec:relWork}

The usage of 2-hop neighborhood knowledge is not new: many broadcast/multicast algorithms have tried to reduce and eliminate redundant transmissions based on this information. Optimized Link State Routing (OLSR) protocol for Mobile Ad Hoc Networks (MANET) \cite{olsr} is one of the many protocols that do so. In OLSR, multipoint relays (MPR) are selected to minimize the number of unnecessary retransmissions that would flood messages in the entire network. Each node selects its MPR set among one-hop neighbors in such a manner that the set covers all nodes that are 2-hop away.  With the MPR method, OLSR can provide efficient routes in terms of number of hops.
  
2-hop neighborhood was also investigated in geographic routing protocols that are probably more related to our proposition. Some real-time algorithms for WSN \cite{thvr, thvr2, path} are based on 2-hop neighborhood knowledge. Some approaches propose to map a packet deadline to a velocity with the key idea of taking routing decisions based on the 2-hop velocity to meet the desired QoS. Another protocol \cite{tpgfplus} uses the 2-hop neighborhood information to find more paths of shorter lengths for duty-cycled systems. 

Recent studies on the performance in $k$-hop neighborhood-based geographic routing, where $k=1,2,3,\ldots$, have established that the improvement from 1-hop searching to 2-hop searching is generally substantial \cite{k-hop}. As the improvement from $k$-hop searching ($k\geqslant2$) to ($k+1$)-hop searching gets smaller due to the fact that the distance between nodes is now shorter, the authors in \cite{adnl} showed that the 2-hop neighborhood knowledge is sufficient to get acceptable results in terms of accuracy in $k$-hop neighborhood-based distributed node localization for WSN. Hence, our motivation for 2-hop neighborhood knowledge in cover set selection. Our work uses 2-hop neighborhood information at the application level in a cross-layer-like fashion to mainly determine which node or set of nodes will be more suitable to relay a large amount of image packets. Under the multi-path assumption, our approach defines metrics to probabilistically determine the likelihood of multi-path transmissions required by a given frame capture rate. Therefore, our approach is very targeted for mission-critical surveillance applications putting clearly the application's criticality in the control loop.

\section{Image sensor node selection method}
\label{sec:selection}

First, as assumed in many sensor-based surveillance applications, each sensor node is aware of its location through  either GPS capability or the ability to estimate their position through anchor nodes that have GPS capability \cite{Tan10}. We also assume that a mobile robot has GPS capability and that it will periodically advertise its position so that sensor nodes within its neighborhood, and possibly at $k$ hops, know the robot's position. If necessary, an adaptive position update mechanism can be implemented \cite{Wang:2009:ALU:1502534.1502545}.

Second, many distributed algorithms in ad-hoc networks implement an initialization phase that usually exchanges HELLO or a similar message to obtain information on one node's 1-hop neighborhood (this phase is usually referred to as the neighbor discovery phase). Extending to 2-hop neighbors can be done quite easy at a relatively low cost as each sensor node can broadcast its neighbor table at the end of the neighbor discovery phase. Specific to our scenario, each sensor during the neighbor discovery phase would collect from their neighbors their node id, GPS position, camera line of sight, angle of view and depth of view of the camera. This list is non-exhaustive and other parameters can be sent at initialization if necessary.

\subsection{2-hop neighborhood information}
\label{sec:2hopIntro}

\begin{figure}[htb]
\begin{center}
\includegraphics[width=.9\linewidth]{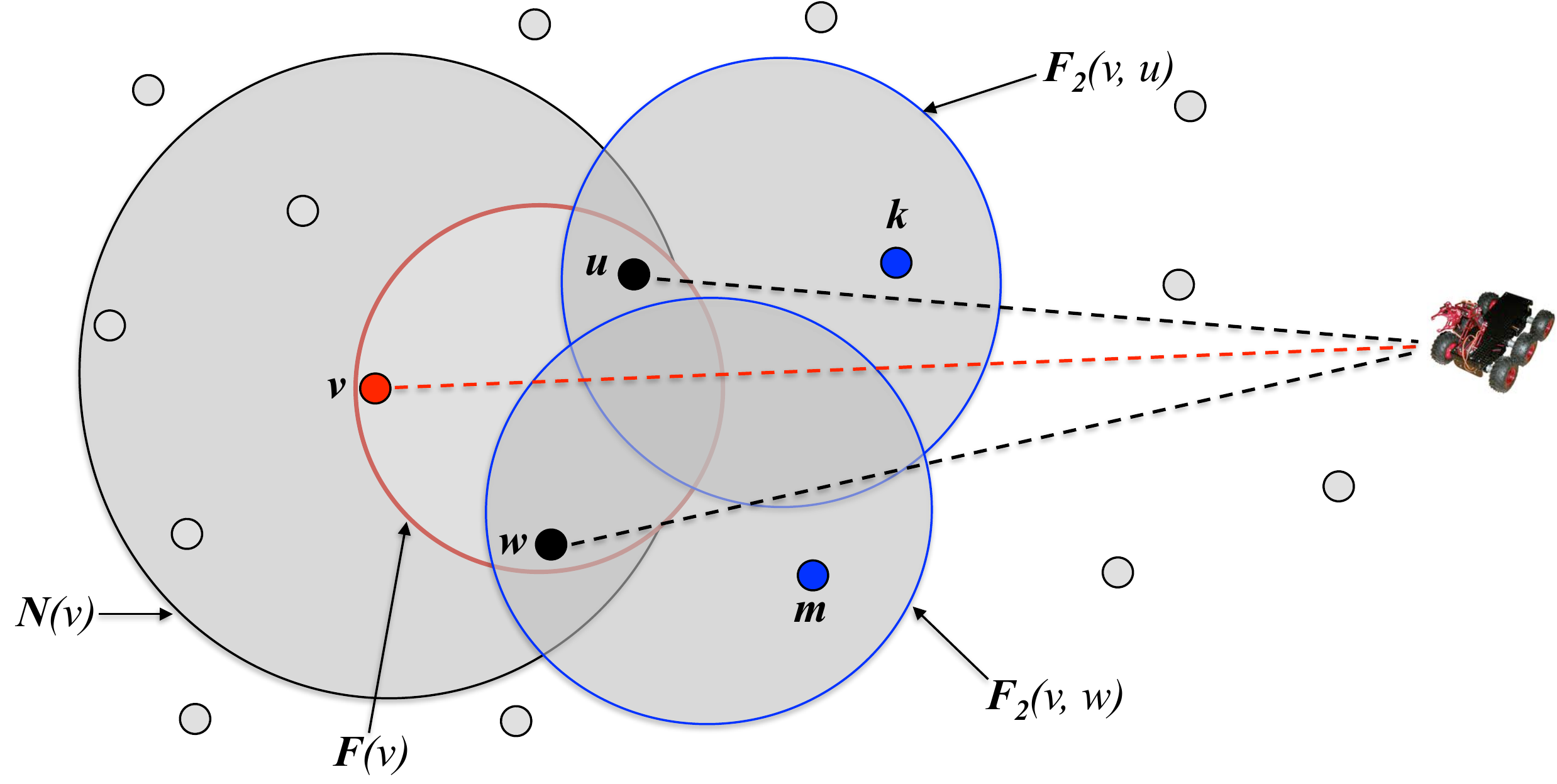}
\caption{Potential 1-hop \& 2-hop forwarders for node $v$}
\label{fig:forwarders}
\end{center}
\end{figure}

Let us denote by $\textbf{N}(v)$ node $v$'s 1-hop neighbor set, see figure \ref{fig:forwarders}.
$\textbf{F}(v)$ is defined as the set of $v$'s 1-hop potential forwarders, i.e. the closest 1-hop neighbors to the robot sink:
\begin{equation*}
	\textbf{F}(v)  = \Big\{u | d(u, Sink) < d(v, Sink), u \in N(v) \Big\}
\end{equation*}
where  $d(u, Sink)$ is the Euclidean distance to the robot sink.
The set of $v$'s 2-hop potential forwarders is denoted $\textbf{F}_\textbf{2}(v)$. Then, the subset of $v$'s 2-hop potential forwarders with node $u$ as intermediate node is defined as follows:
\begin{equation*}
	\textbf{F}_\textbf{2}(v, u)   = \Big\{k | d(k, Sink) < d(u, Sink),  u \in F(v), k \in N(u) \Big\}
\end{equation*}

\subsection{Cover Set Selection Approach}
\label{sec:selectionMethod} 
Mission-critical applications have QoS requirements such as reliability of received data at the sink with strict delay, especially for visual information. Congestion and contention on the radio medium are the main source of packet losses as the network load increases. Therefore the capture rate of image sensor nodes should guide the choice of cover sets as it will have a high impact on the data transmission performance, both at MAC and network level. Multi-path routing is often regarded as a solution to improve communication performance in WSN: data transmission reliability, bandwidth aggregation, load balanced transmission, congestion-free transmission, low latency transmission, \dots The establishment of multiple paths between a pair (source, destination) for data transmission can increase reliability and some approaches such as \cite{reInForm} even use the path redundancy to send multiple copies of the same packet on the various paths to the Sink. In our proposition this is not the technic we adopt. We use multiple paths for both load balanced and congestion-free transmissions when a large amount of visual information need to be sent on the network. Therefore, the idea we develop here is to link the image capture rate (or the number of images to send) to the need of multiple paths: the higher the capture rate of a node, the higher is the need for multiple paths towards the Sink.

Through the usage of the 2-hop neighborhood guided by the capture rate we define a first metric for cover sets selection: $R_ {2-hop}$ measures the likelihood of a given cover set to find as many needed 2-hop paths as required by the capture rate. $R_ {2-hop}$ for a given cover set $Co_i(v)$ of node $v$ is given by the equation below:
\begin{equation*}
 \label{eq:2hopRatio}
R_ {2-hop} (Co_i(v)) = \frac{1}{|Co_i(v)|} \sum_{w=1}^{|Co_i(v)|}
    \frac
       {\qquad
              |F_2(w)|
       \qquad}{NbOptimalPaths(w) }
\end{equation*}
where $|F_2(w)|$ is the number of $w$'s 2-hop potential forwarders, $w \in Co_i(v))$, and $NbOptimalPaths(w)$ is the number of optimal paths of $w$. We define $NbOptimalPaths(w)$ to be proportional to $w$'s image capture rate. Linking the capture rate to the number of required paths to correctly transfer images is an original feature of our approach because capture rates can be very different from an image sensor to another since some geographical areas could be at a higher criticality level than others \cite{Pham11a}. As scheduling of sensors is very dynamic for these mission-critical applications, the best cover set is highly dependent on the required capture rate. 

Alone, the $R_ {2-hop}$ metric does not necessarily guarantee improved performance for establishing disjointed paths. For a given cover set, having enough 2-hop potential forwarders, i.e. $R_ {2-hop}$ is high, is important but these 2-hop potential forwarders may have few relay nodes themselves, i.e. 1-hop potential forwarders, and may also share most of them making disjointed paths for decreasing inter-path interferences very difficult or impossible. A cover set with many unshared relay nodes per 2-hop forwarder has better efficiency to set up disjointed paths for load balancing purposes. Therefore a second criterion, noted $R_ {relay}$, is combined with $R_ {2-hop}$ as follows:
\begin{equation*}
 \label{eq:relayRatio}
 R_ {relay}(Co_i(v)) = \frac{1}{|Co_i(v)|} \sum_{w=1}^{|Co_i(v)|}
    \frac
       {\qquad
              |F(w)|
       \qquad}{|F_2(w)| }
\end{equation*} 
where $|F(w)|$ and $|F_2(w)|$ are the number of $w$'s 1-hop and 2-hop potential forwarders respectively, $w \in Co_i(v))$. The ratio $\frac{|F(w)|}{|F_2(w)|}$ expresses the likelihood that a 2-hop forwarder have several unshared relay nodes. For example, let $w$ be a cover set member with 3 2-hop forwarders. If the number of unshared relay neighbors is also 3, this ratio is 1 and there is potentially for each 2-hop neighbor a different relay node. If this ratio exceeds 1, it is even better. However, there is no strict guarantees since a single 2-hop neighbor may well have all the relay nodes. Here we made a trade-off between the difficulty and to overhead to obtain and consider very accurate information and this is the reason why we propose a probabilistic approach that has the advantage of being very simple and requiring only a small additional cost in terms of message exchanged compared to traditional 1-hop information. The method we take here is an on-demand method: as all nodes know their 2-hop neighbors, a node $v$ with cover sets would send a request to its cover set members to get their list of 2-hop neighbors.

Each cover set is then associated to a Transmission Quality ($TQ$) value which is used to score and classify cover sets at a sentry node. TQ is computed based on previous metrics with weights to indicate the importance degree of each metric according to equation below: 
 \begin{equation*}
 \label{eq:score}
TQ(Co_i(v)) = \alpha \times R_ {2-hop}(Co_i(v))  + \beta \times  R_ {relay}(Co_i(v))
\end{equation*}
where $\alpha + \beta = 1$. 
For a given sentry node, the cover set with the highest $TQ$ value has better performance for transmitting image packets, i.e. with low latency and less packet losses. The selection algorithm can also consider the remaining energy of cover sets which can be defined as the minimum energy of the cover set members. Now, to be consistent with our proposed selection method, a multi-path extension of GPSR will be described in the next section to ensure that routing decisions are also based on the 2-hop neighborhood information that has been taken for image transmission at the application level.

\section{GPSR Extension}
\label{sec:tgpsr}

\subsection{Greedy Perimeter Stateless Routing (GPSR)}
GPRS  is a geographic routing protocol originally designed for MANETS which has been rapidly adapted for WSN \cite{gpsr,agem}. Each node is aware of its location and of its 1-hop neighbors' locations. GPSR has two strategies for forwarding data packets to the destination: \textit{Greedy Forwarding} and \textit{Perimeter Forwarding}. In Greedy Forwarding, whenever a node needs to forward a data packet, it chooses the closest neighbor to the destination as the next hop. The packet will be transmitted and relayed hop-by-hop by choosing at each hop the next neighbor node which is the closest by Euclidian distance to the destination.

Sometimes, the greedy forwarding strategy fails to find a neighbor closer to the destination than itself because of voids or holes due to random deployment, obstacles that obstruct radio signals or node failures. To overcome this problem, Perimeter Forwarding is used to route packets around voids using the right-hand rule: packets will move around the void until it reaches a node closer to the destination than the node which has initiated the Perimeter Forwarding process. To reach the final destination, a Greedy Forwarding phase is then started from this point.

\subsection{T-GPSR: a 2-hop-information-based GPSR Extension}
The T-GPSR extension is essentially based on collecting 1-hop \& 2-hop neighborhood information during both the neighbor discovery process and the cover set selection process performed at the application level. This is similar to some so-called {\em cross-layer} approaches where information from lower levels are used by higher levels. These approaches are widely used in sensor networks but it is necessary to pay particular attention to what information should be considered and to avoid those that are difficult to get in a network with a large number of nodes. For example, network and/or link load is a difficult information to estimate, especially in a wireless network where the size of buffer queues is not simply correlated with the network load due to interference phenomena or contention on the radio support.

As an extension to GPSR, our proposed routing scheme incorporates an additional strategy, called \textit{2-Hop-based Greedy Forwarding}, for taking account the 2-hop neighborhood information. In \textit{2-Hop-based Greedy Forwarding}, whenever a source node $v$ needs to forward a data packet, it chooses the closest 2-hop potential forwarder to the final destination in $F_2(v)$. Thus, packets are sent to this 2-hop potential forwarder as the temporary destination through one of $v$'s 1-hop potential forwarder, in $F(v)$, acting as relay node. For instance, if we look back at figure \ref{fig:forwarders}, source node $v$ selects the 2-hop potential forwarder $m$ as temporary destination and 1-hop  potential forwarder $w$ as relay. When a relay node receives a data packet to forward, there is no additional next hop discovery to execute: it will just send the packet to the associated temporary destination, $m$ in this case. Therefore forwarding decisions  occur only every two hops which contributes to decrease latency especially when an important number of hops is required to reach the robot sink. On the other hand, a temporary destination that receives a data packet to forward behaves as a source node. This process is repeatedly executed until the data packet reaches the robot sink. This strategy is prone to failure if $|F_2(v)|=0$, i.e. $v$ has no 2-hop potential forwarder. In this case, T-GPSR will adopt the original GPSR \textit{Greedy Forwarding} mode on $F(v)$. Finally, GPSR \textit{Perimeter Forwarding} is used when the greedy forwarding fails. 

\section{Simulation results}
\label{sec:results}

We evaluate our proposal with the OMNET++/Castalia framework (http://castalia.research.nicta.com.au). We consider an homogenous wireless image sensor network where 400 image sensor nodes are randomly deployed in a $2km*2km$ area.  Sensor nodes have an $60^o$ angle of view, a depth of view of $125m$ and a communication range of $150m$. On this network topology, we perform a set of simulations to show the benefit of our cover set selection approach. In what follows, we consider three scenarios for transmitting images:
\begin{itemize}
\item Scenario 1:  no selection algorithm is required. For instance, each sentry selects the first active cover set in its cover set table. The routing layer uses GPSR.
\item Scenario 2: our selection mechanism is performed at the application level, and GPSR is also used for routing.
\item Scenario 3: our selection mechanism is again performed at the application level but now T-GPSR is used at the routing layer.
\end{itemize}
In scenarios 2 and 3, the routing layer uses additional information from the selection algorithm such as shared relay nodes for example. In all scenarios, CSMA/CA is used at the MAC layer and the radio link throughput is 250kbps. We monitored the average packet loss rate, the average quality of received images at the robot sink and the average image transmission delay to the robot sink. Given that the time scale of data transmission is much smaller that the robot's velocity, the impact of the robot sink mobility is negligible. 

As described in Section \ref{sec:intro}, when  a node $v$ detects an event such as an intrusion, it will \textit{(a)} send one or several images to the robot sink and  \textit{(b)} activate one of its cover sets. On activation, cover set members will also send one or several images to the robot sink to provide more information for disambiguation purposes. The simulation model implements the transmission of real image files by taking into account all communication layers. We use an optimized image format for sensor networks that combines robustness with respect to packet losses, low power consumption in compression and small file size with a selectable quality factor \cite{Dur11,Lec12}. In addition, image packets can be received in any order at the robot sink which is a desirable feature with multi-path routing. In our case, an image has $320*320$ pixels with 256 gray levels for a raw size of 102400B. We then use a quality factor of 50 that gives a final image size of 16621B. By setting the maximum payload size to 90B, the encoding scheme gives 205 packets.
\subsection{Packet loss}

\begin{figure}[H]
\begin{center}
\includegraphics[width=1.0\linewidth]{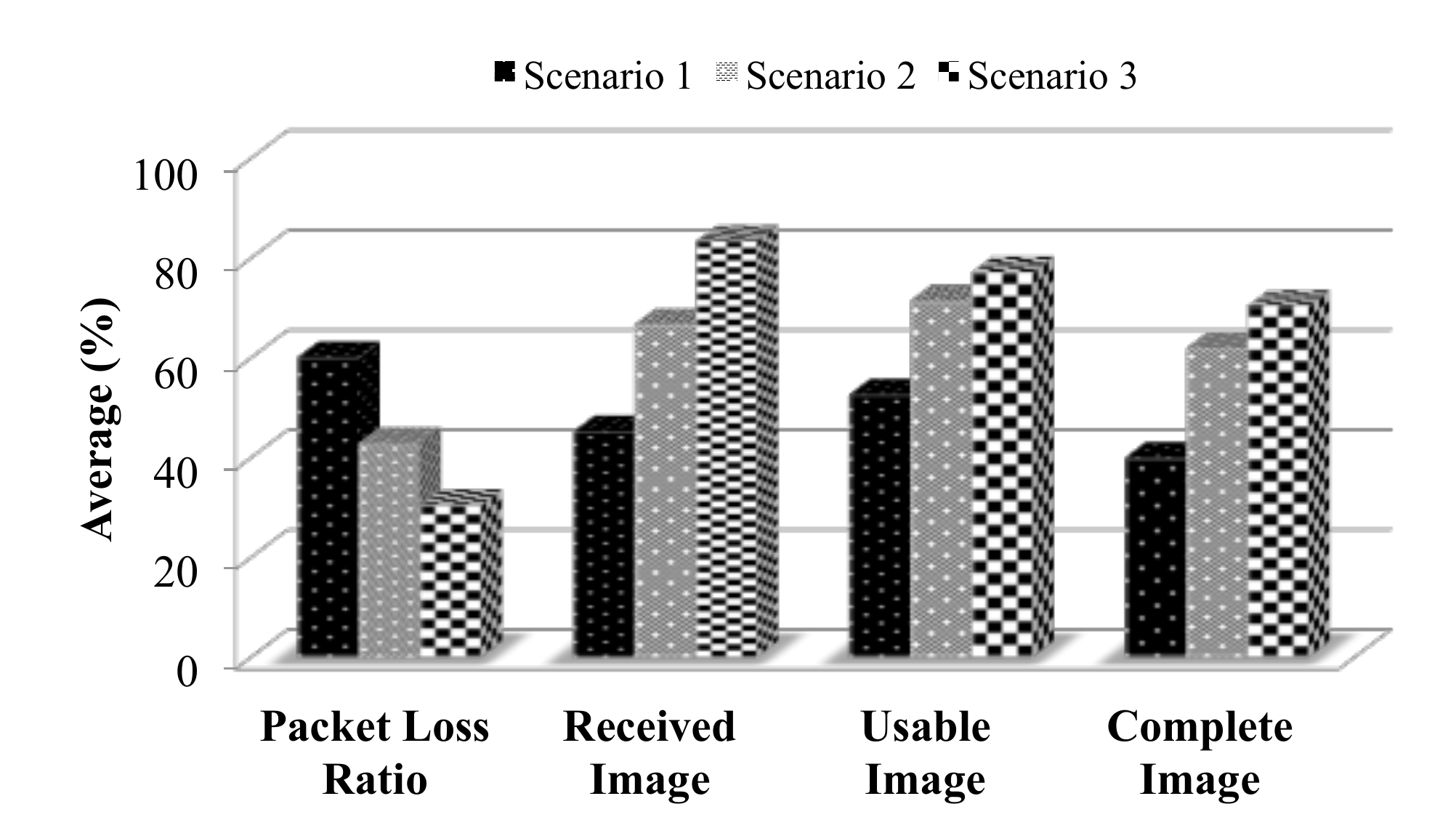}
\caption{Received mage statistics}
\label{fig:imageStatistic}
\end{center}
\end{figure}

As shown in figure \ref{fig:imageStatistic}, in scenarios 2 and 3 the average loss rate does not exceed 40\% compared to Scenario 1. Scenario 3 shows a smaller loss rate than Scenario 2 thanks to the 2-hop neighborhood knowledge of T-GPSR which increases reliability. However, when the image capture rate gets higher, the numerous simultaneous transmissions of images create congestion and inter-path interferences to name a few issues. 

In figure \ref{fig:imageStatistic}, we can see that the percentage of received images of scenarios 2 and 3 is much higher compared to Scenario 1. This result shows that the cover set selection mechanism succeeds in reducing contention in image transmission. In addition, the percentage of received images in Scenario 3 is larger by 20\% than Scenario 2 clearly showing the additional benefit of T-GPSR 2-hop information usage.

\subsection{Image quality}

\begin{figure}[H]
\begin{center}
\includegraphics[width=1.0\linewidth]{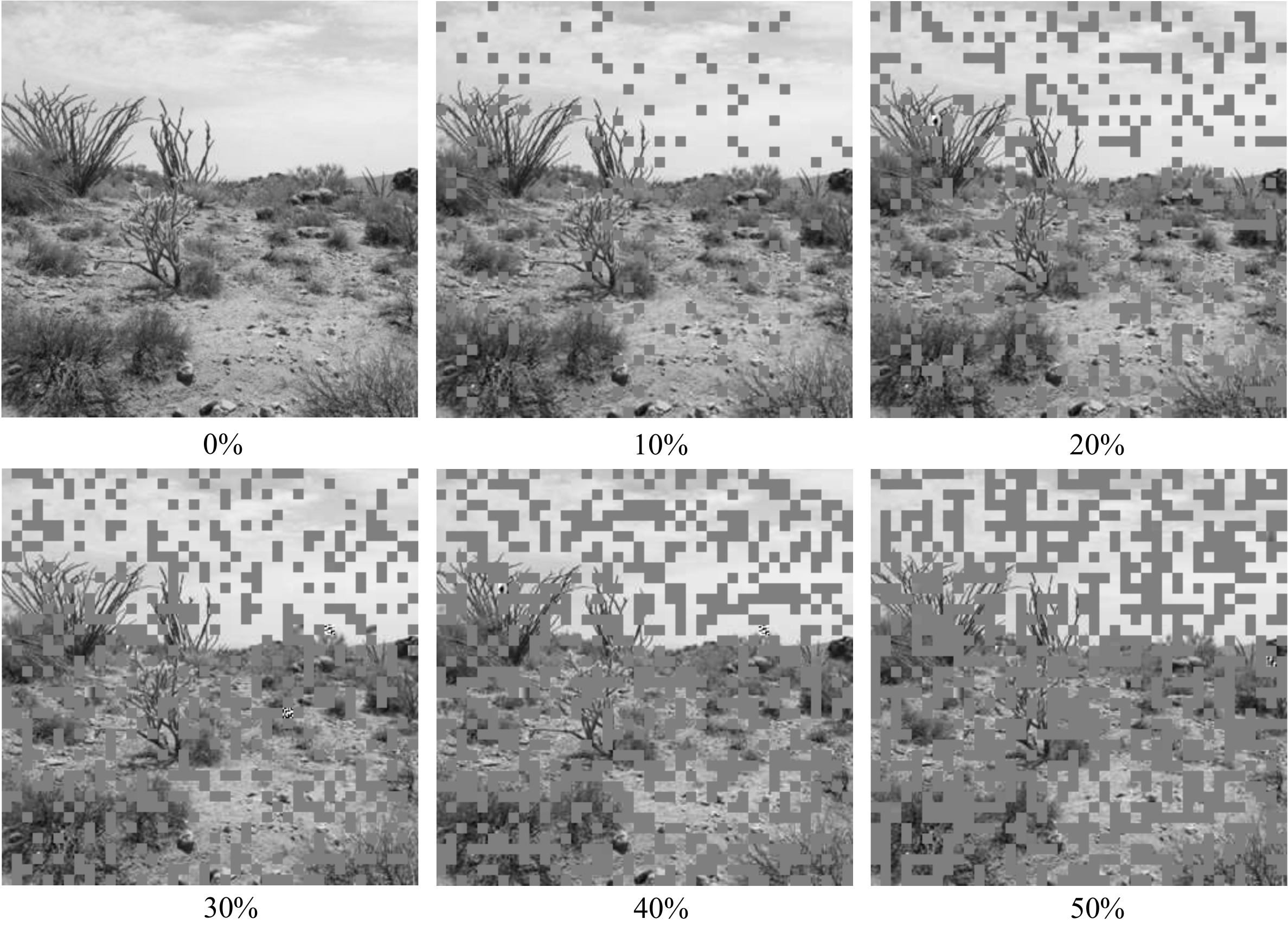}
\includegraphics[width=1.0\linewidth]{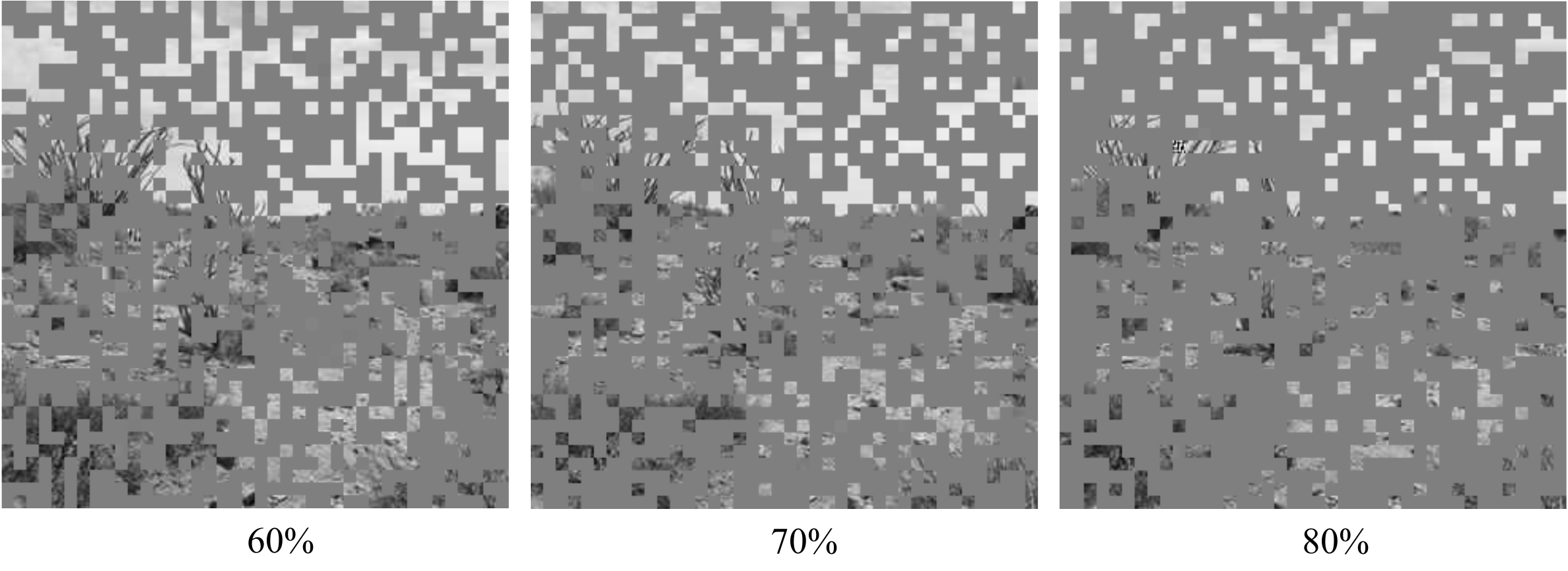}
\caption{Image quality at the robot sink at various packet loss ratios}
\label{fig:qualitylevel}
\end{center}
\end{figure}

In the context of a mission-critical application, detecting events is important but receiving high quality images is also very important. Reception of a large number of images at the robot sink does not necessarily mean that they are all exploitable. The packet loss ratio has a direct impact on the received image quality, and in all our simulations we observed that an image with more than $60\%$ of packet losses is visually not exploitable (for identification purposes for instance). Also, a received image is either complete (no packet loss) or truncated. Figure \ref{fig:qualitylevel} shows the $320*320$ original image and received images with various packet loss ratios.

Although image quality may be very application-dependent, we decide to set the threshold at 60\% of packet losses and we will classify an image as unusable when the packet loss ratio is greater than 60\%. By opposition, when the packet loss ratio is smaller than 60\% the image will be classified as usable. 
With this convention, figure \ref{fig:imageStatistic} shows that our selection approach (scenarios 2 and 3) increases the number of usable images at the robot sink compared to Scenario 1. In addition, most of these usable images in these scenarios have complete (no packet loss). Once again, in Scenario 3, we can see that the usage of T-GPSR to reflect at the routing layer the 2-hop information collected at the application level further reduces packet losses, thus increasing the image quality at the  robot sink.

\subsection{Image reception latency}
\begin{figure}[htbp]
\begin{center}
\includegraphics[width=1.0\linewidth]{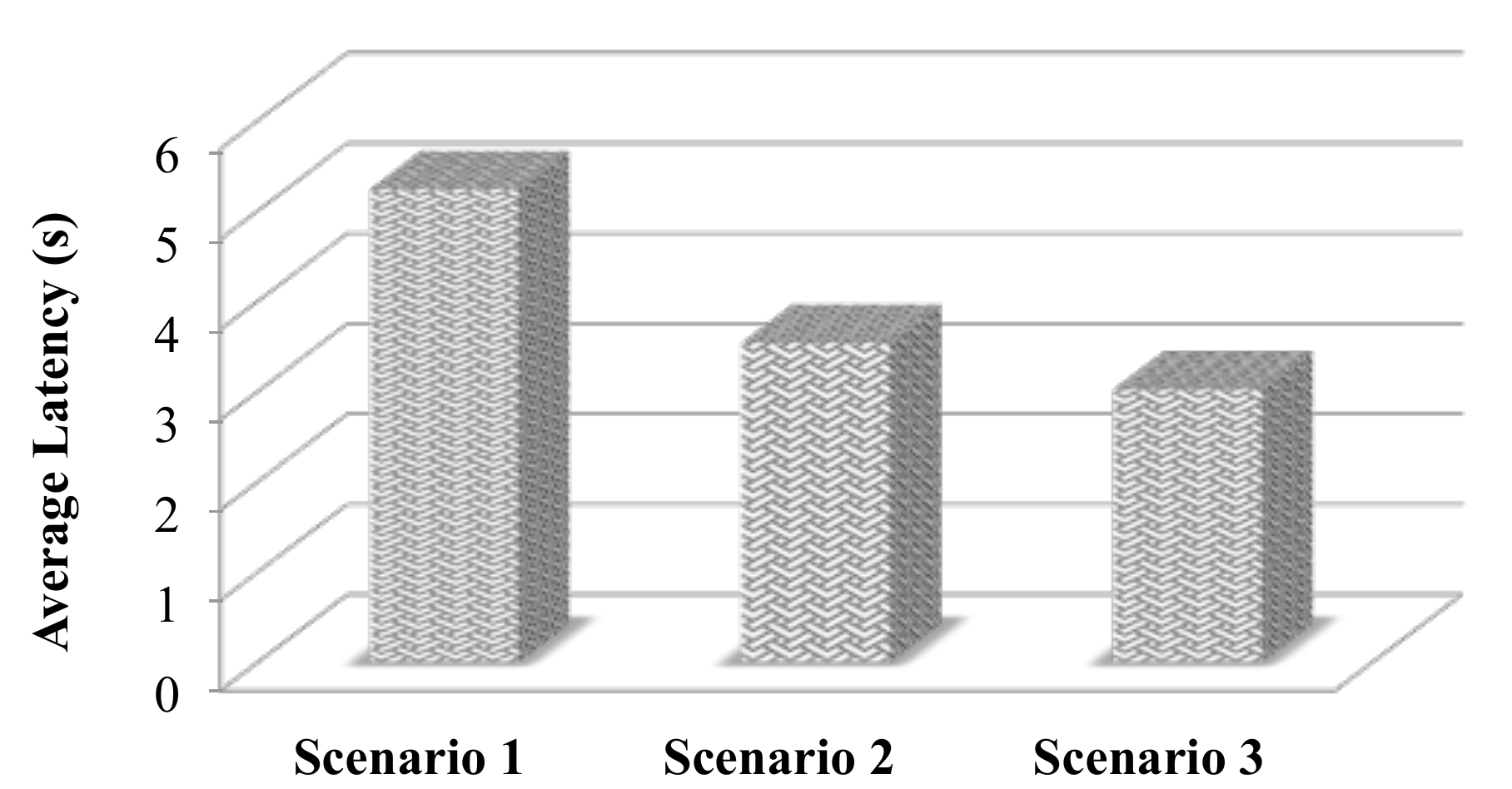}
\caption{Average image reception latency ratio}
\label{fig:latency}
\end{center}
\end{figure}

As stated previously, achieving the lowest latency for image reception at the robot sink is also very important. The packet loss rate can have a strong impact on the image reception latency. The implemented decoder can display an image regardless of the number of received packets and regardless of their reception order. However, we still need a timer that is set at the reception of the first image packet and that will trigger the display of the image regardless of the number of packet actually received. When the number of lost packets is high, the latency can be as high as the display timer which is set to 10s. With low loss probability, the latency is much lower and depends on the number of hops. Although neither the API various transmission limitations nor hardware limitations are accurately modeled, we can however compare the latency achieved by our approach with the case when there is no 2-hop neighborhood information used. In our current simulation model, a single image can be received in 0.94s in the very best case. Figure \ref{fig:latency} compares the reception average delay of the three scenarios. 

\section{Conclusion}
\label{sec:conclusion}

In this paper, we considered Robot-assisted Search\&Rescue operations enhanced with some fixed image sensor nodes capable of capturing and sending visual information to a robot sink. Image sensor nodes can have very basic event detection capabilities while mobile robots have more powerful processing capabilities to act as a Sink for the fixed image sensor nodes. We first proposed an optimized image node selection approach based on 2-hop neighborhood information to determine the most relevant cover sets to be activated. The motivation is to increase reliability for image transmission by reducing both funneling effect and contention on the medium. Then, in order to be consistent with the proposed selection approach, a multi-path extension of Greedy Perimeter Stateless Routing (called T-GPSR) where routing decisions are also based on 2-hop neighborhood information has been proposed. One key point of our proposition is to link and consider several important parameters that depend on the image capture rate and the network topology. 

Simulations were carried out to show the benefits of our proposition. We simulated event detection systems where images are sent for disambiguation purposes. Performance evaluations have shown that our proposal reduces the packet loss ratio to provide better received image quality at the robot sink. Our approach is particularly efficient when the amount of data is large, which is the case with increasing image capture rates or image sensor node density.

\bibliographystyle{IEEEtran} \bibliography{references,reference-WD13-1,reference-WD13-2}

\begin{thebibliography}{10}
\providecommand{\url}[1]{#1}
\csname url@samestyle\endcsname
\providecommand{\newblock}{\relax}
\providecommand{\bibinfo}[2]{#2}
\providecommand{\BIBentrySTDinterwordspacing}{\spaceskip=0pt\relax}
\providecommand{\BIBentryALTinterwordstretchfactor}{4}
\providecommand{\BIBentryALTinterwordspacing}{\spaceskip=\fontdimen2\font plus
\BIBentryALTinterwordstretchfactor\fontdimen3\font minus
  \fontdimen4\font\relax}
\providecommand{\BIBforeignlanguage}[2]{{%
\expandafter\ifx\csname l@#1\endcsname\relax
\typeout{** WARNING: IEEEtran.bst: No hyphenation pattern has been}%
\typeout{** loaded for the language `#1'. Using the pattern for}%
\typeout{** the default language instead.}%
\else
\language=\csname l@#1\endcsname
\fi
#2}}
\providecommand{\BIBdecl}{\relax}
\BIBdecl

\bibitem{Mak10}
A.~Makhoul, R.~Saadi, and C.~Pham, ``Risk management in intrusion detection
  applications with wireless video sensor networks,'' in \emph{IEEE WCNC},
  2010.

\bibitem{Pham11a}
C.~Pham, A.~Makhoul, and R.~Saadi, ``Risk-based adaptive scheduling in randomly
  deployed video sensor networks for critical surveillance applications,''
  \emph{Journal of Network and Computer Applications}, vol.~34, pp. 783--795,
  2011.

\bibitem{Pha13-wimob}
C.~Pham, V.~Lecuire, and J.-M. Moureaux, ``Performances of multi-hops image
  transmissions on ieee 802.15.4 wireless sensor networks for surveillance
  applications,'' in \emph{IEEE WiMob}, 2013.

\bibitem{olsr}
T.~Clausen and P.~Jaquet, ``Optimized link state routing protocol,'' \emph{in
  IETF MANET, RFC 3626}, October 2003.

\bibitem{thvr}
Y.~Li, C.~S. Chen, Y.~Q. Song, Z.~Wang, and Y.~Sun, ``A two-hop based real-time
  routing protocol in wireless sensor networks,'' \emph{Proc. IEEE WFCS}, vol.
  5, no. 2, p. 65–74, May. 2008.

\bibitem{thvr2}
------, ``Enhancing realtime delivery in wireless sensor networks with two-hop
  information,'' \emph{IEEE Transactions on Industrial Informatics}, vol. 5,
  no.2, p. 113–122, May. 2009.

\bibitem{path}
P.~Rezayat, M.~Mahdavi, M.~Ghasemzadeh, and M.~Sarram, ``A novel real-time
  power aware routing protocol in wireless sensor networks,'' \emph{IJCSNS},
  vol. 10, no.4, p. 300–305, April 2010.

\bibitem{tpgfplus}
Y.~Dong, L.~Shu, G.~Han, M.~Guizani, and T.~Hara, ``Geographic multipath
  routing in duty-cycled wireless sensor networks: One-hop or two-hop?''
  \emph{MobiCom 2011}, 2011.

\bibitem{k-hop}
C.~S. Chen, Y.~Li, and Y.~Q. Song, ``An exploration of geographic routing with
  k-hop based searching in wireless sensor networks,'' \emph{ChinaCom'08}, p.
  376–381, Aug. 2008.

\bibitem{adnl}
J.~Champ and V.~Boudet, ``Adnl-angle: accurate distributed node localization
  for wireless sensor networks with angle of arrival information,'' \emph{Proc.
  ADHOC-NOW'10}, pp. 177--190, 2010.

\bibitem{Tan10}
S.~Tanvir, E.~Schiller, and A.~Duda, ``Propagation protocols for network-wide
  localization based on two-way ranging,'' \emph{in IEEE WCNC}, 2010.

\bibitem{Wang:2009:ALU:1502534.1502545}
\BIBentryALTinterwordspacing
G.~Wang, T.~Wang, W.~Jia, M.~Guo, and J.~Li, ``Adaptive location updates for
  mobile sinks in wireless sensor networks,'' \emph{J. Supercomput.}, vol.~47,
  no.~2, pp. 127--145, Feb. 2009. [Online]. Available:
  \url{http://dx.doi.org/10.1007/s11227-008-0181-5}
\BIBentrySTDinterwordspacing

\bibitem{reInForm}
B.~Deb, S.~Bhatnagar, and B.~Nath, ``Reinform: Reliable information forwarding
  using multiples paths in sensor networks,'' \emph{Proc. IEEE ICLCN}, pp.
  406--415, 2003.

\bibitem{gpsr}
{B. Karp, H. T. Kung}, ``Gpsr: Greedy perimeter stateless routing for wireless
  networks,'' \emph{ACM, MobiCom}, 2000.

\bibitem{agem}
S.~Medjiah, T.~Ahmed, and F.~Krief, ``Agem: Adaptative greedy-compass
  energy-aware multipath routing protocol for wmsns,'' \emph{IEEE CCNC}, 2010.

\bibitem{Dur11}
C.~Duran-Faundez, V.~Lecuire, and F.~Lepage, ``Tiny block-size coding for
  energy-efficient image compression and communication in wireless camera
  sensor networks,'' \emph{Signal Processing: Image Communication}, vol.~26,
  pp. 466--481, 2011.

\bibitem{Lec12}
V.~Lecuire, L.~Makkaoui, and J.-M. Moureaux, ``Fast zonal dct for energy
  conservation in wireless image sensor networks,'' \emph{Electronics Letters},
  vol.~48, 2012.

\end{thebibliography}

\end{document}